\documentclass{article}

\usepackage{iclr2024_conference,times}
\iclrfinalcopy

\usepackage[utf8]{inputenc} % allow utf-8 input
\usepackage[T1]{fontenc}    % use 8-bit T1 fonts
\usepackage{hyperref}       % hyperlinks
\usepackage{url}            % simple URL typesetting
\usepackage{booktabs}       % professional-quality tables
\usepackage{amsfonts}       % blackboard math symbols
\usepackage{nicefrac}       % compact symbols for 1/2, etc.
\usepackage{microtype}      % microtypography

\usepackage{graphicx}
\usepackage{enumitem}
\usepackage[page]{appendix}
\usepackage{multirow}
\hypersetup{colorlinks=true}
\usepackage{color}
\usepackage{xcolor}

\newcommand\textgreen[1]{\textcolor[rgb]{0,0.5,0}{\textbf{#1}}}
\newcommand\textred[1]{\textcolor[rgb]{1,0,0}{\textbf{#1}}}
\newcommand{\footnoteurl}[1]{\footnote{\url{#1}}}

\title{Highly Efficient Real-Time Streaming and Fully On-Device Speaker Diarization with Multi-Stage Clustering}

\author{%
  Quan Wang$^\star$ \quad Yiling Huang$^\star$ \quad Han Lu \quad Guanlong Zhao \quad Ignacio Lopez Moreno \\
  Google LLC, USA \qquad $^\star$Equal contribution\\
  \texttt{\{quanw,yilinghuang,luha,guanlongzhao,elnota\}@google.com} \\
}

\fancypagestyle{firstpage}{
  \lhead{\begin{picture}(0,0)\put(0,-3){\includegraphics[width=0.16\linewidth]{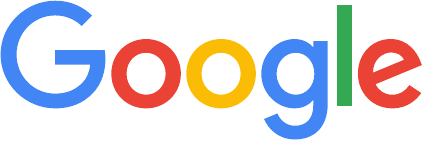}}\end{picture}}
}

\begin{document}

\maketitle
\thispagestyle{firstpage}
\begin{abstract}
While recent research advances in speaker diarization mostly focus on improving the quality of diarization results, there is also an increasing interest in improving the efficiency of diarization systems. In this paper, we demonstrate that a multi-stage clustering strategy that uses different clustering algorithms for input of different lengths can address multi-faceted challenges of on-device speaker diarization applications. Specifically, a fallback clusterer is used to handle short-form inputs; a main clusterer is used to handle medium-length inputs; and a pre-clusterer is used to compress long-form inputs before they are processed by the main clusterer. Both the main clusterer and the pre-clusterer can be configured with an upper bound of the computational complexity to adapt to devices with different resource constraints. This multi-stage clustering strategy is critical for streaming on-device speaker diarization systems, where the budgets of CPU, memory and battery are tight.
\end{abstract}

\section{Introduction}
\label{sec:intro}
Research advances in the speaker diarization community have been tackling various challenges in the past decade. Different clustering algorithms including agglomerative hierarchical clustering (AHC)~\cite{garcia2017speaker}, K-means~\cite{dimitriadis2017developing} and spectral clustering~\cite{wang2017speaker} had been explored. Specifically, many works had been proposed to further improve the spectral clustering algorithms for speaker diarization, such as auto-tune~\cite{park2019auto}, speaker turn constraints~\cite{xia2022turn}, and multi-scale segmentation~\cite{park2021multi}.
To better leverage training datasets that are annotated with timestamped speaker labels, supervised diarization approaches have been explored, including UIS-RNN~\cite{zhang2019fully}, DNC~\cite{li2019discriminative}, and EEND~\cite{fujita2019end,e2ediarizationpatent}. 
Approaches such as TS-VAD~\cite{medennikov2020target} and EEND-SS~\cite{ueda2022eend} had been proposed to solve the speech separation and diarization problems jointly~\cite{von2019all,chazan2018attention}.
Various other advances are described and discussed in literature reviews and tutorials~\cite{park2021review,zhang2022odysseytutorial}.

Despite these advances, another challenge that prevents speaker diarization systems from being widely used in production environments is the efficiency of the system, which is also a relatively less discussed topic in the speaker diarization community. In this paper, we are particularly interested in streaming on-device speaker diarization for mobile phones, such as annotating speaker labels in a recording session~\cite{speakerlabelsblog}. The requirements for such applications are multi-faceted:

\begin{enumerate}
    \item The diarization system must perform well on audio data from multiple domains. Since supervised diarization algorithms such as UIS-RNN~\cite{zhang2019fully} and EEND~\cite{fujita2019end} are highly dependent on the domain of the training data, and often suffer from insufficient training data, we prefer to use unsupervised clustering algorithms in such applications.
    \item The diarization system must perform well on input audio of variable lengths, from a few seconds to a few hours. 
    \item The system must work in a streaming fashion, producing real-time speaker labels while audio being recorded by the microphone on the device. 
    \item The diarization system must be optimized to be highly efficient, to work within the limited budget of CPU, memory and power. Particularly, the computational cost of the clustering algorithm must be upper bounded to avoid out-of-memory (OOM) or excessive battery drain issues on mobile devices, even if the input audio is hours long.
\end{enumerate}

To meet these requirements, we propose a multi-stage clustering strategy that combines AHC and spectral clustering, and leverages the strength of both algorithms. Based on our experiments, with a pre-define threshold, AHC is good at identifying single speaker versus multiple speakers for short-form audio, but often ends up with too many small clusters, especially for long-form speech. Spectral clustering is great at estimating the number of speakers with the eigen-gap approach, but usually under the assumptions that there are at least two different speakers, and there are sufficient data points to be clustered. At the same time, the computational cost of spectral clustering is too expensive for long-form speech, mostly due to the calculation and eigen decomposition of the Laplacian matrix.
Based on the above observations, we use different clustering algorithms for inputs of different lengths. When input audio is short, we use AHC to avoid the insufficient data problem of spectral clustering; when input audio is of medium length, we use spectral clustering for better speaker count estimate, and eliminate dependency on a pre-defined AHC threshold; when input audio is long, we first use AHC as a pre-clusterer to compress the inputs to hierarchical cluster centroids, then use spectral clustering to further cluster these centroids. We enforce an upper bound on the number of inputs to the AHC pre-clusterer by caching and re-using previous AHC centroids, such that the overall computational cost of the entire diarization system is always bounded.

\section{Baseline system}
\label{sec:baseline}

Our speaker diarization system is largely built on top of the Turn-to-Diarize system described in~\cite{xia2022turn}, which consists of a speaker turn detection model, a speaker encoder model, and unsupervised clustering.

\subsection{Feature frontend}
We used a shared feature frontend for the speaker turn detection model and the speaker encoder model. This frontend first applies automatic gain control~\cite{prabhavalkar2015automatic} to the input audio, then extracts 32ms-long Hanning-windowed frames with a step of 10ms. For each frame, 128-dimensional log Mel-filterbank energies (LFBE) are computed in the range between 125Hz and 7500Hz. These filterbank energies are then stacked by 4 frames and subsampled by 3 frames, resulting in final features of 512 dimensions with a frame rate of 30ms. These features are then filtered by a CLDNN based Voice Activity Detection (VAD) model~\cite{zazo2016feature} before fed into the speaker turn detection and the speaker encoder models.

\subsection{Speaker turn detection}
\subsubsection{Model architecture}
The speaker turn detection model is a Transformer Transducer (T-T)~\cite{zhang2020transformer} trained to output automatic speech recognition (ASR) transcripts augmented with a special token \texttt{<st>} to represent a speaker turn.
The Transformer Transducer architecture includes an audio encoder, a label encoder, and a joint network that produces the final output distribution over all possible tokens.

The audio encoder has 15 layers of Transformer blocks. Each block has 32 left context and no right context. The hyper-parameters for each repeated block can be found in Table~\ref{table:scd-arch}. We also use a stacking layer after the second transformer block to change the frame rate from 30ms to 90ms,  and an unstacking layer after the 13th transformer block to change the frame rate from 90ms back to 30ms, to speed up the audio encoder as proposed in~\cite{tripathi2020transformer}.

We use a LSTM-based label encoder that has a single layer of 128 dimensions.

For the joint network, we have a projection layer that projects the audio encoder output to 256 dimensions. At the output of the joint network, it produces a distribution over 75 possible graphemes\footnoteurl{https://github.com/google/speaker-id/blob/master/publications/Multi-stage/graphemes.syms} with a softmax layer. For optimization, we follow the same hyper-parameters described in~\cite{zhang2020transformer}.

This T-T model has $\sim$ 13M parameters in total, and is trained with the token-level loss that is introduced in~\cite{zhao2022augmenting}, with these hyper-parameters of the loss function: weight for word errors $\alpha=1$; weight for speaker turn false accepts $\beta=100$; weight for speaker turn false rejects $\gamma=100$; weight for RNN-T loss $\lambda=0.03$; and weight for customized minimum edit distance alignment $k=1.1$.

\begin{table}[t]
    \centering
    \caption{Hyper-parameters of a Transformer block for the audio encoder of the speaker turn detection model.}
    \begin{tabular}{c|c}
    \toprule
    Input feature projection &  256 \\
    Dense layer 1 & 1024 \\
    Dense layer 2 & 256 \\
    Number attention heads & 8 \\
    Head dimension & 64  \\ 
    Dropout ratio & 0.1  \\
    \bottomrule
    \end{tabular}
    \label{table:scd-arch}
\end{table}

\subsubsection{Training data}
The training data for the speaker turn detection model include Fisher~\cite{cieri2004fisher} training subset, Callhome American English~\cite{canavan1997callhome} training subset, AMI training subset, ICSI training subset \footnoteurl{https://github.com/kaldi-asr/kaldi/tree/master/egs/icsi}, 4545 hours of internal long-form videos, and 80 hours of internal simulated business meeting recordings. 
\subsection{Speaker encoder}
\subsubsection{Model architecture}
\begin{figure*}
    % Source image: https://docs.google.com/drawings/d/1WQZJEFT9oO-PWWt2Ss6Q9MEdC3mXJThAzbvFg4djkuE/edit
	\centering
	\includegraphics[width=0.7\linewidth]{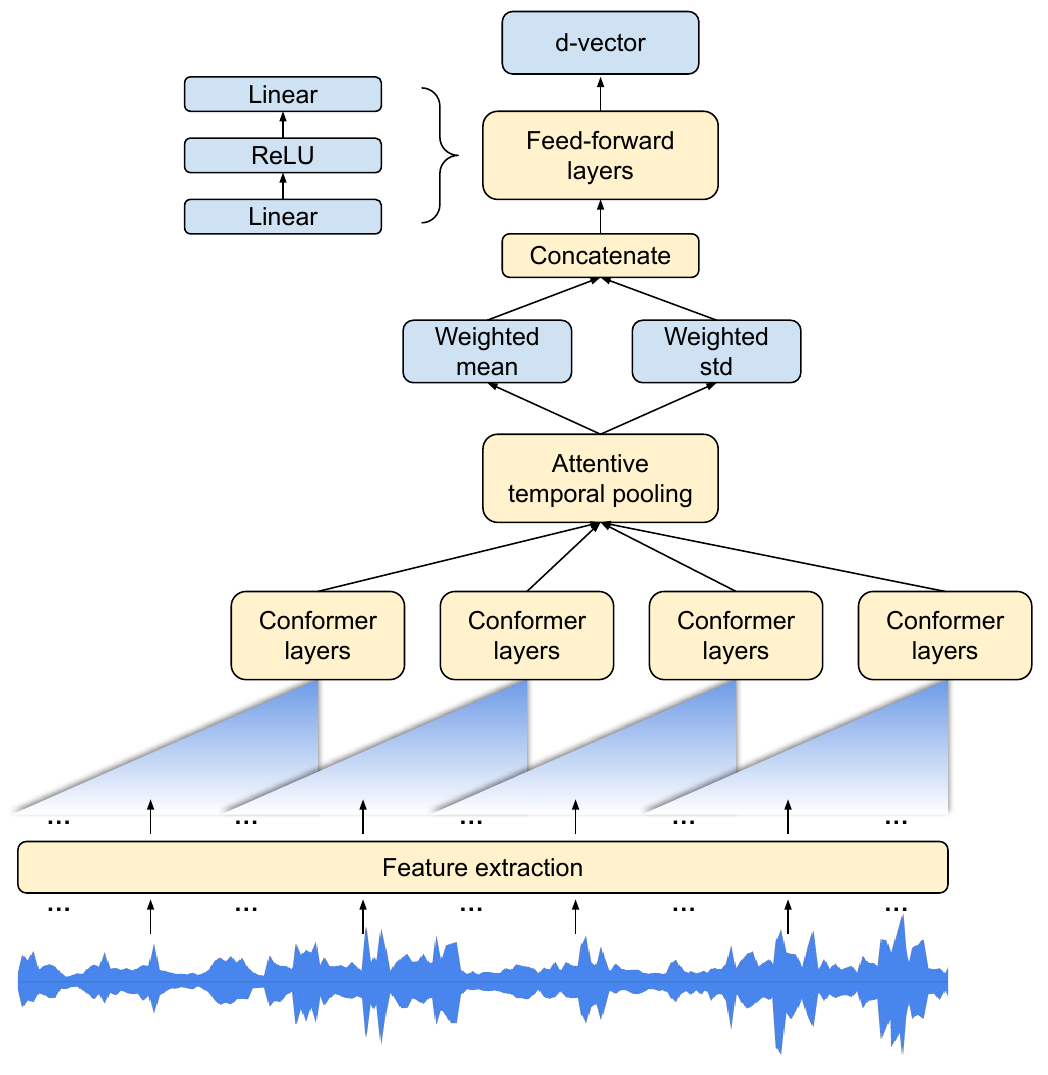}
	\caption{Architecture of the speaker encoder model.}
	\label{fig:speaker_encoder}
\end{figure*}

The speaker encoder is a text-independent speaker recognition model trained with the generalized end-to-end extended-set softmax (GE2E-XS) loss~\cite{ge2e,pelecanos2021dr}, which consists of 12 conformer~\cite{gulati2020conformer} layers each of 144-dim, followed by an attentive temporal pooling module~\cite{wang2022attentive}. The speaker encoder model has $\sim$ 7.4M parameters in total, and the architecture is illustrated in Fig.~\ref{fig:speaker_encoder}.

To reduce the length of the sequence to be clustered, the speaker encoder produces only one embedding (\emph{i.e.} d-vector) for each speaker turn, or every 6 seconds if a single turn is longer than that. To achieve great performance and streaming diarization at the same time, we run spectral clustering in an online fashion: every time when we have a new speaker embedding, we run spectral clustering on the entire sequence of all existing embeddings. This means it's possible to correct previously predicted speaker labels in a later clustering step.

\subsubsection{Training data}
The training data for the speaker encoder model are the same as the data used in ~\cite{pelecanos2022parameter} and \cite{koizumi2023miipher}. 
It is a mixture~\cite{chojnacka2021speakerstew} consisting mostly of vendor collected speech queries from different language varieties\footnote{72 languages in the training data include: Afrikaans, Akan, Albanian, Arabic, Assamese, Basque, Bengali, Bulgarian, Cantonese, Catalan, Croatian, Czech, Danish, Dutch, English, Estonian, Filipino, Finnish, French, Galician, German, Greek, Gujarati, Haitian, Hausa, Hebrew, Hindi, Hungarian, Icelandic, Igbo, Indonesian, Italian, Japanese, Kannada, Kazakh, Kinyarwanda, Korean, Lithuanian, Macedonian, Malagasy, Malay, Malayalam, Mandarin Chinese (Simplified), Mandarin Chinese (Traditional), Marathi, Mongolian, Norwegian, Odia, Oromo, Polish, Portuguese, Punjabi, Romanian, Russian, Samoan, Serbian, Sesotho, Sindhi, Slovak, Spanish, Swedish, Tamil, Telugu, Thai, Tibetan, Turkish, Ukrainian, Urdu, Uzbek, Vietnamese, Yoruba, Zulu.} using devices such as laptops and cell phones, as well as public training datasets including LibriVox~\cite{librivox_2020_1}, CN-Celeb~\cite{fan_2020_1} and LDC sourced data (\emph{i.e.} Fisher~\cite{Cieri2004_1}, Mixer 4/5~\cite{Brandschain2020_1}, and TIMIT~\cite{garofolo_1993_1}). The language distribution is shown in Table~\ref{tab:sidtraindata}.

During training, we apply data augmentation techniques based on noise and room simulation effects~\cite{lippmann1987multi,ko2017study,kim2017generation}. Similar augmentation techniques~\cite{garcia-romero_2012_1,lei_2012_1,avila_2014_1,snyder_2018_1,huang_2019_1} were previously used for speaker recognition. Noise is added to the training utterances with an SNR ranging from 3dB to 15dB. The signal amplitude is also varied from a scale of 0.01x to 1.2x.

\begin{table}
\centering
    \label{tab:sidtraindata} 
  \caption{Training data of the speaker encoder model across languages. The first 9 rows correspond to the major language varieties of vendor collected data, while the last row includes both vendor collected data of minor languages, and public datasets such as LibriVox, CN-Celeb, and LDC sourced data. All numbers are indicated in thousands (as indicated by [k]).}
\begin{tabular}{l c c }
\hline
    Data source      & Number of speakers [k]  & Number of utterances [k] \\ \hline
  Vendor data: English (India)              & 6.0 & 2900  \\
  Vendor data: English (US)               & 46.7 & 3329 \\
  Vendor data: French              & 5.8 & 2458  \\ 
  Vendor data: Hindi              & 8.3 & 1642 \\
  Vendor data: Italian              & 5.2 & 2251 \\
  Vendor data: Japanese              & 6.2 & 2048  \\
  Vendor data: Korean               & 5.4 & 1407  \\
  Vendor data: Portuguese (Brazil)   & 5.8 & 1675  \\
  Vendor data: Spanish                   & 4.6 & 2189  \\
  Public datasets + Vendor data: other languages & 139.8 & 54499    \\ \hline
\end{tabular}
\end{table}

\subsection{Constraints of the baseline system}
This baseline system has shown great potential in many applications, but still have several constraints when deploying to mobile applications:
\begin{enumerate}
    \item Spectral clustering uses the eigen-gap criterion to estimate the number of speakers. But this approach usually assumes there are at least two speakers. It often fails to distinguish single speaker versus multiple speakers.
    \item Spectral clustering works best when there are sufficient inputs to be clustered. This means for the first few minutes during the streaming diarization, the quality of the streaming outputs can be significantly lower, resulting in low user retention.
    \item The computational cost of spectral clustering on $N$ embeddings is $O(N^\omega)$, where $\omega$ depends on the specific implementation and the desired precision~\cite{pan1999complexity}. Similar to matrix multiplication, $\omega$ has a theoretical lower bound of $\omega \geq 2.37$ ~\cite{coppersmith1987matrix}. The high cost will result in huge latency when the input is long-form, \emph{e.g.} when user attempts to diarize hours-long recordings, thus is not acceptable for streaming applications.
\end{enumerate}

\section{Multi-stage clustering}

\begin{figure*}
    % Source image: https://docs.google.com/drawings/d/1NTu9UO8-vMeeis2LBPUBjinsX0vxjBzXfZWNlY3Bwo8/edit
	\centering
	\includegraphics[width=0.99\linewidth]{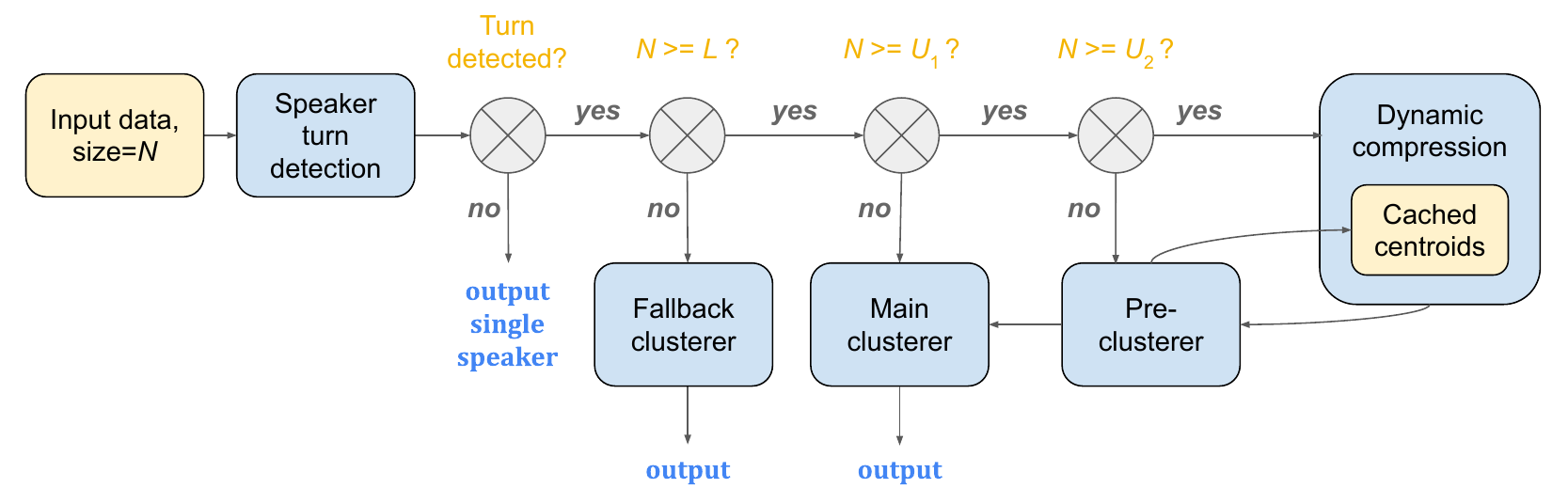}
	\caption{Diagram of the multi-stage clustering strategy. $L$ and $U_1$ are the lower bound and upper bound of the main clusterer, respectively. $U_2$ is the upper bound of the pre-clusterer.}
	\label{fig:diagram}
\end{figure*}

A high-level diagram of the multi-stage clustering strategy is illustrated in Fig.~\ref{fig:diagram}. Assume the input sequence consists of $N$ speaker embeddings from the speaker encoder model. The clustering algorithm being used will vary based on different values of $N$, as described below.

\subsection{Speaker turn based decision}
The speaker encoder model will produce a new speaker embedding both at detected speaker turns and when reaching a maximal length of a segment (\emph{i.e.} 6 seconds). If no speaker turn token \texttt{<st>} was detected for the $N$ speaker embeddings, we directly output a single speaker without running any clustering algorithm.

\subsection{Fallback clusterer}
Since spectral clustering cannot handle small inputs very well, we could use a different fallback clustering algorithm when $N<L$, where $L$ is the lower bound of the number of inputs to spectral clustering. In this paper, we use AHC with the average linkage type as our fallback clusterer. The stopping criterion of the AHC is based on a  threshold pre-determined on dev datasets.

\subsection{Main clusterer}

If the input size $N$ is between the lower and upper bounds of the spectral clustering algorithm, \emph{i.e.} $L \leq N < U_1$, the main clusterer (\emph{i.e.} spectral clustering) is used to cluster the inputs directly.
With this upper bound, the computational cost of spectral clustering will not exceed $O(U_1^{\omega})$~\cite{pan1999complexity}.
Since auto-tune~\cite{park2019auto,xia2022turn} is used to improve the quality of spectral clustering results, the computational cost needs to be multiplied by the number of search steps.
% Besides, previous work observed that auto-tune~\cite{park2019auto} could significantly improve the quality of spectral clustering results, especially when the input length varies in a big range in the evaluation dataset~\cite{xia2022turn}. But with the lower and upper bounds, the benefit of auto-tune is marginal, thus we can remove auto-tune to further save computational cost. \Yiling{update this, since autotune is used in the experiments.}

\subsection{Pre-clusterer}

If $U_1 \leq N < U_2$, we use a pre-clusterer to reduce the size of the clustering inputs first. Here we use AHC with the complete linkage (\emph{a.k.a.} farthest neighbor) type as our pre-clusterer. When the number of clusters reduces to $U_1$, the AHC pre-clusterer will stop, and we compute the centroids of these $U_1$ pre-clusters. Then we feed these $U_1$ centroids as the input to the main clusterer to get the final speaker labels of all original inputs.

\subsection{Dynamic compression}
With the pre-clusterer, we upper bounded the cost of spectral clustering to $O(U_1^{\omega})$. However, the AHC pre-clusterer itself has a computational cost of $O(N^2)$ or larger, depending on implementation. When the input audio is very long, this cost will still be unacceptable. Thus we propose a dynamic compression approach to also enforce an upper bound $U_2$ for the AHC pre-clusterer.

During the streaming diarization process, when $N=U_2$, we trigger the AHC pre-clusterer, and compress the $U_2$ inputs to $U_1$ pre-cluster centroids. We store these $U_1$ centroids and the mapping between these centroids to original inputs in an internal cache. For future clustering steps when $N>U_2$, the first $U_2$ inputs are replaced by the cached $U_1$ centroids, and the AHC pre-clusterer will only run on $N'=U_1+(N-U_2)$ inputs.

% This above step will repeat until $N=2 \cdot U_2-U_1$, when the total number of inputs (previously compressed and new inputs) to the AHC clusterer reaches $U_2$ again. Then we need to update the cache of the $U_1$ centroids, as well as the mapping between these centroids to the $2 \cdot U_2-U_1$ original inputs.

As $N$ increases, $N'$ will hit the upper bound $U_2$ again. Thus we need to update the cached centroids and the centroid-to-input mapping every $U_2-U_1$ steps. The number of cached centroids is always $U_1$, and it maps to $U_2 + (K-1) \cdot (U_2-U_1)$ original inputs, where $K$ is the number of times that the input to AHC clusterer reached $U_2$ previously. For each individual step, the cost of AHC pre-clusterer is upper bounded to $O(U_2^2)$.

\subsubsection{Cost analysis}
With the mutli-stage clustering strategy described above, the input size to the spectral clustering algorithm is upper bounded to $U_1$, and the input size to the AHC algorithm is upper bounded to $U_2$. Thus the overall computational cost of an individual clustering step is upper bounded to $O(U_1^{\omega})+O(U_2^2)$, where $2.37 \leq \omega \leq 3$ depends on the specific implementation. At the same time, due to the dynamic compression, we only need to store at most $U_2$ embeddings/centroids in memory, instead of all the previous $N$ embeddings. Thus both the time complexity and space complexity of the clustering algorithm are upper bounded to a configurable constant.

\section{Experiment settings}
\label{sec:exp}

\subsection{Quality metrics}
\label{sec:metrics}
To evaluate the impact of different components that we proposed in the multi-stage clustering strategy, we perform a careful ablation study, and report the Diarization Error Rate (DER) of different setups on various evaluation datasets.
When computing DER, we tolerate a collar value of 250ms before and after each reference speaker segment boundary.

For AMI, Callhome, Fisher, and ICSI, we convert their customized annotations into the Rich Transcription Time Marked (RTTM) format. DIHARD1 is already using the RTTM format and therefore no conversion is required. We then use the non-overlapping union of the RTTM segments as the final Un-partitioned Evaluation Map (UEM), except for DIHARD1 which provides its own UEM files.

Due to the scale of the evaluations, we use an internal MapReduce~\cite{dean2008mapreduce} based C++ implementation to calculate the Diarization Error Rate (DER) reported in Section~\ref{sec:results}. Thus the DER numbers reported in this paper may have some discrepancies with numbers computed with other libraries such as pyannote.metrics~\cite{bredin2017pyannote}. 

\subsection{Efficiency metrics}
To estimate the long-form efficiency, we report the total number of floating point operations (FLOPs) required to perform one individual clustering step for a specific value of $N$. 

The FLOPs estimation (\emph{e.g.} in Table~\ref{tab:long_results}) is based on PyPAPI\footnoteurl{https://flozz.github.io/pypapi} by counting \texttt{PAPI\_FP\_OPS}, using the AHC implementation from scikit-learn\footnoteurl{https://scikit-learn.org/stable/modules/generated/sklearn.cluster.AgglomerativeClustering.html} and Python spectral clustering from~\cite{wang2017speaker}\footnoteurl{https://github.com/wq2012/SpectralCluster}. Note that the FLOPs of the feature frontend and neural network models are excluded in Table~\ref{tab:long_results} as they are constant.

The original script that we used to estimate FLOPs of multi-stage clustering and the FLOPs results for more values of $N$ are open sourced on GitHub\footnoteurl{https://github.com/google/speaker-id/tree/master/publications/Multi-stage/flops}.

\subsection{Evaluation data}
Our evaluation datasets are listed in Table~\ref{table:datasets}. Some additional details are listed below:
\begin{enumerate}
    \item For AMI~\cite{carletta2005ami}, we use the official ``Full-corpus-ASR partition"\footnoteurl{https://groups.inf.ed.ac.uk/ami/corpus/datasets.shtml} test subset. The speaker label ground truth was obtained based on the word-level annotations in the v1.6.2 AMI manual annotation release.
    \item For Callhome American English Speech~\cite{canavan1997callhome}, we only evaluate on the official evaluation subset, as the training subset is used for training the speaker turn detection model.
    \item For DIHARD1~\cite{ryant2018first}, we use the eval subsets from 9 sources, but remove all YouTube-based data. We use the original UEM files from the challenge to make our results comparable to Track 2 results\footnoteurl{https://dihardchallenge.github.io/dihard1/results.html\#track2}. Since the original UEM include overlapping speech which is not handled by our system, we expect the DER numbers to be high on this dataset.
    \item For Fisher~\cite{cieri2004fisher}, we use a test subset of 172 utterances.\footnoteurl{https://github.com/google/speaker-id/blob/master/publications/Multi-stage/evaluation/Fisher/eval_whitelist.txt}
    \item For ICSI~\cite{janin2003icsi}, we use a test subset that is segmented to shorter utterances of less than 30 min\footnoteurl{https://github.com/google/speaker-id/tree/master/publications/Multi-stage/evaluation/ICSI/eval_segmentation.csv}.
    \item ``Outbound" and ``Inbound" are vendor-provided call center telephone conversations between call center attendants and customers, initiated by the call center and by customers, respectively. These had also been used in~\cite{xia2022turn}.
\end{enumerate}

\begin{table*}
    \centering
    \caption{A list of our evaluation datasets. Some of the datasets had been filtered or processed due to license issues.}
    \begin{tabular}{c|c|c|c|c|c}
    \toprule
    \multirow{2}{*}{Dataset} & \multirow{2}{*}{Domain} & Num.  & Num. & Avg. & Avg.  \\
    & & utt & hours & spk/utt & length\\ \hline
    AMI & Meeting & 16 & 9.1 & 4 & 34min \\ 
    Callhome & Telephone & 20 & 1.7 & 2 & 5min \\
    DIHARD1 & Mixed & 114 & 16.2 & 3.3 & 8.5min \\
    Fisher & Telephone & 172 & 28.7 & 2 & 10min \\
    ICSI & Meeting & 13 & 5 & 6.4 & 23min \\
    Inbound & Telephone & 250 & 21 & 4.3 & 5min \\
    Outbound & Telephone & 450 & 45.6 & 2 & 6.1min \\ \bottomrule 
    \end{tabular}
    \label{table:datasets}
\end{table*}

As discussed in Sec~\ref{sec:intro}, short-form performance is also critical for on-device streaming applications. To measure the quality of short-form inputs, we created segmented versions of the above mentioned datasets, where each utterance is approximately only 30, 60 or 120 seconds.

When preparing the datasets for evaluation, we notice that different data sources have different representation of speaker label annotations. For example, the AMI dataset annotation is almost on a word-level basis, where most of the RTTM segments are extremely short. As mentioned in Section~\ref{sec:metrics}, during the evaluation, we extrude a collar value of 250ms centered around each reference speaker segment boundary. This leads to potential excessive removal of the audio from the original UEM. In order to mitigate this issue, we include an additional step in our data processing pipeline which merges neighboring RTTM segments that are from the same speaker and are close to each other. For our experiments, we specifically consider any gap smaller than 0.01 second to be close enough for RTTM merging.

% Our experiments abide by Google AI principles~\cite{aiprinciples}.

\section{Experimental results}
\label{sec:results}
\subsection{Short-form results}
\label{sec:shortform-results}
First, the DER results on short-form segmented versions of the evaluation datasets are shown in Table~\ref{tab:short_results}. We can clearly see that the DER on all of the 30s, 60s, and 120s segmented datasets are pretty bad when there is no fallback clusterer, \emph{i.e.} only main clusterer is used. By introducing the AHC fallback clusterer, the DER is significantly reduced on these short-form datasets. Specifically, with a larger $L$, the fallback clusterer is capable to handle longer audio. This validates our hypothesis that spectral clustering does not work well given insufficient input data points.

On the other hand, we still prefer spectral clustering as our main clusterer. As shown in Table~\ref{tab:speaker_diff}, with threshold tuning, the AHC clusterer can be configured to produce decent DER across our testsets, but inevitably has the tendency to overestimate the number of clusters. This gets much worse when the audio is longer (\emph{e.g.} AMI and ICSI both have unacceptably big speaker count errors). As a resolution, we find $L=50$ to be a sweet point between short-form and long-form quality, which uses AHC fallback clusterer long enough while yielding not-too-far-off number of speakers.

\begin{table*}
  \centering
  \caption{Speaker count errors and breakdown. We show the Mean Absolute Error (MAE) of speaker count for each dataset, along with the breakdown numbers in the parentheses: (percentage of utterances with correct number of speakers, percentage of utterances with overestimated number of speakers, percentage of utterances with underestimated number of speakers).
  %$L=0$ indicates we only use the main clusterer, while $L=\inf$ indicates only the fallback clusterer is used.
  }
%   \scriptsize
    \small
    \hspace*{-3.5cm}
    \begin{tabular}{c|ccccccc|c}
    \toprule
    \multicolumn{1}{c|}{} & \multicolumn{8}{c}{Mean absolute error of speaker count (\% of correct estimate,\ \% of over estimate,\ \% of under estimate)} \\
    System & AMI & Callhome & DIHARD1 & Fisher & ICSI & Inbound & Outbound & Average \\ \hline
    Main clusterer & 0.94 (19,50,31) & 0.2 (80,15,5) & 1.43 (32,33,35) & 0.05 (95,5,0) & 2.38 (15,8,77) & 1.68 (20,9,71) & 0.34 (72,28,0) & \textgreen{1.0 (48,21,31)} \\  % spectral
    Fallback clusterer & 12.19 (0,100,0) & 0.85 (30,60,10) & 3.94 (4,89,7) & 2.46 (3,97,0) & 11.62 (0,100,0) & 2.22 (8,87,5) & 3.37 (3,97,0) & \textred{5.24 (7,90,3)} \\  % AHC
    \bottomrule
    \end{tabular}
  \label{tab:speaker_diff}
\end{table*}

\begin{table*}
  \centering
  \caption{DER (\%) on short-form segmented versions of the evaluation datasets for different setups of the fallback clusterer. $L$ is the lower bound of the main clusterer.}
    \begin{tabular}{c|c|cccc}
    \toprule
    \multirow{2}{*}{Datasets} & \multirow{2}{*}{Length} &  \multicolumn{3}{c}{DER (\%)}  \\
     &  & No fallback & $L=25$ & $L=50$ \\ \hline
    AMI & \multirow{8}{*}{30s} & 13.25 & 6.73 & 6.64 \\
    Callhome &  & 12.34 & 6.8 & 6.8 \\
    DIHARD1  &  & 33.04	& 26.47 & 26.39 \\
    Fisher &  & 12.84 &  4.08 & 4.08 \\
    ICSI &  & 17.96 &  5.35 & 5.35 \\
    Inbound &  & 13.8 & 5.36 & 4.96 \\
    Outbound  &  & 17.91 & 9.39 & 9.22 \\ 
    \bf{Average}  &  & \textred{17.31} & 9.17 & \textgreen{9.06} \\ \hline
    AMI & \multirow{8}{*}{60s} & 16.81 & 8.48 & 6.95 \\
    Callhome &  & 12.11 & 5.74 & 6.06 \\
    DIHARD1  &  & 35.85 & 26.9 & 27.03 \\
    Fisher &  & 8.6 & 3.59 & 3.25 \\
    ICSI &  & 20.99 & 7.24 & 6.16 \\
    Inbound &  & 13.76 & 6.02 & 5.1 \\
    Outbound  &  & 14.48 & 9.38 & 8.66 \\ 
    \bf{Average}  &  & \textred{17.51} & 9.62 & \textgreen{9.03} \\ \hline
    AMI & \multirow{8}{*}{120s} & 14.74 & 14.06 & 8.04 \\
    Callhome &  & 6.56 & 6.16 & 6.44 \\
    DIHARD1  &  & 33.34 & 32.27 & 27.19 \\
    Fisher &  & 2.43 & 2.41 & 2.54 \\
    ICSI &  & 13.35 & 12.98 & 7.06 \\
    Inbound &  & 10.02 & 9.22 & 5.49 \\
    Outbound  &  & 7.55 & 7.36 & 7.94 \\ 
    \bf{Average}  &  & \textred{12.57} & 12.07 & \textgreen{9.24} \\
    \bottomrule
    \end{tabular}
  \label{tab:short_results}
\end{table*}

\subsection{Long-form results}
Table~\ref{tab:long_results} shows the diarization evaluation results from different multi-stage clustering setups on various test datasets. For the system (1) where no spectral clustering is used and AHC is directly used to produce final results, we find the performance is significantly worse than a system like (2) or (3) with $L=0$ or $L=50$ on datasets with only 2 speakers (\emph{i.e.} Callhome, Fisher and Outbound). This is because AHC tends to over-estimate the number of clusters, and spectral clustering is better at finding the correct number of clusters via the eigen-gap criterion. This echoes the observation from Sec~\ref{sec:shortform-results} regarding overestimation.

With the introduction of pre-clustering and dynamic compression, experiments (3)vs(4)vs(5), (4)vs(6) and (7)vs(8) present expected degradation in diarization quality. Specifically, the degradation is more significant when we use a small value for $U_2$, and more significant on AMI and ICSI datasets, as the utterances in these datasets are much longer in average than other datasets. As a trade-off, the long-form computational cost is remarkably reduced when we set an finite number of $U_1$ and $U_2$, according to the FLOPs of the clustering when $N=2000$ (corresponding to $\sim 2$ hours of audio).

\begin{table*}
  \centering
  \caption{DER (\%) on different evaluation datasets for different setups of multi-stage clustering. We also report FLOPs of performing one individual clustering step when $N=2000$. 
  % For the main clusterer, $L=\inf$ indicates we do not use spectral clustering at all, but directly use AHC to produce final results; $U_1=\inf$ indicates we do not use AHC pre-clusterer.
  }
  % I have already re-ordered the rows. So, $L=50$, $U_1=100$, $U_2=600$ is our best trade-off.
%   \scriptsize
    \small
    \hspace*{-3cm}
    \begin{tabular}{c|ccc|ccccccc|c|c}
    \toprule
     & \multicolumn{3}{c|}{System} & \multicolumn{8}{c|}{DER (\%)} & 
    FLOPs  \\
    Exp & Fallback & Main clusterer & Pre-clusterer  & AMI & Callhome & DIHARD1 & Fisher & ICSI & Inbound & Outbound & Average & $N=2000$ \\ \hline
    1 & Yes & - & - & 6.09 & \textred{6.8} & 29.22 & \textred{2.54} & 7.18 & 5.76 & \textred{9.26} & 9.55 & $2.6 \times 10^7$ \\  % fallback AHC
    2 & - & Yes & - & 7.81 & 2.71 & 34.09 & 1.58 & 12.41 & 9.77 & 6.83 & 10.74 & \textred{$7.7 \times 10^9$} \\  % spectral
    3 & $L=50$ & $U_1=\inf$ & - & 7.8 & 2.71 & 34.33 & 1.58 & 11.98 & 9.19 & 6.98 & 10.65 & \textred{$7.7 \times 10^9$} \\  % AHC, spectral
    4 & $L=50$ & $U_1=300$ & $U_2=\inf$ & 9.88 & 2.7 & 34.27 & 1.58 & 11.12 & 9.2 & 6.93 & 10.81 & $2.1 \times 10^8$ \\  % AHC, AHC + spectral + dynamic compression
    5 & $L=50$ & $U_1=100$ & $U_2=\inf$ & 15.29 & 2.69 & 33.12 & 1.57 & 18.37 & 9.79 & 5.98 & 12.40 & $6.4 \times 10^7$ \\  % AHC, AHC + spectral
    % $L=50$ & $U_1=50$ & $U_2=\inf$ & 10.08 & 2.99 & 33.09 & 1.87 & 17.03 & 9.57 & 6.19 & $3.0 \times 10^7$ \\  % AHC, AHC + spectral
    6 & $L=50$ & $U_1=300$ & $U_2=600$ & 17.55 & 2.7 & 34.24 & 1.58 & 13.1 & 9.13 & 6.89 & 12.17 & $1.8 \times 10^8$ \\  % AHC, AHC + spectral + dynamic compression
    7 & $L=50$ & $U_1=100$ & $U_2=600$ & 18.01 & 2.7 & 33.38 & 1.57 & 17.97 & 9.8 & 5.98 & 12.77 & $3.9 \times 10^7$ \\  % AHC, AHC + spectral + dynamic compression
    8 & $L=50$ & $U_1=100$ & $U_2=300$ & 24.94 & 2.7 & 33.24 & 1.57 & 30.79 & 9.59 & 6.22 & 15.58 & $3.5 \times 10^7$ \\  % AHC, AHC + spectral + dynamic compression
    \bottomrule
    \end{tabular}
  \label{tab:long_results}
\end{table*}

\subsection{On-device CPU runtime benchmarking}

\begin{figure}
    % Source image: https://docs.google.com/presentation/d/1qghTs-uLAgn9pT583cn9EOMvsi-BH_FaLopPFOc8gJ8/edit?resourcekey=0-61QVgxsVhyg4G3nH--mNzQ#slide=id.g16e6318ad4e_1_228
	\centering
	\includegraphics[width=0.7\linewidth]{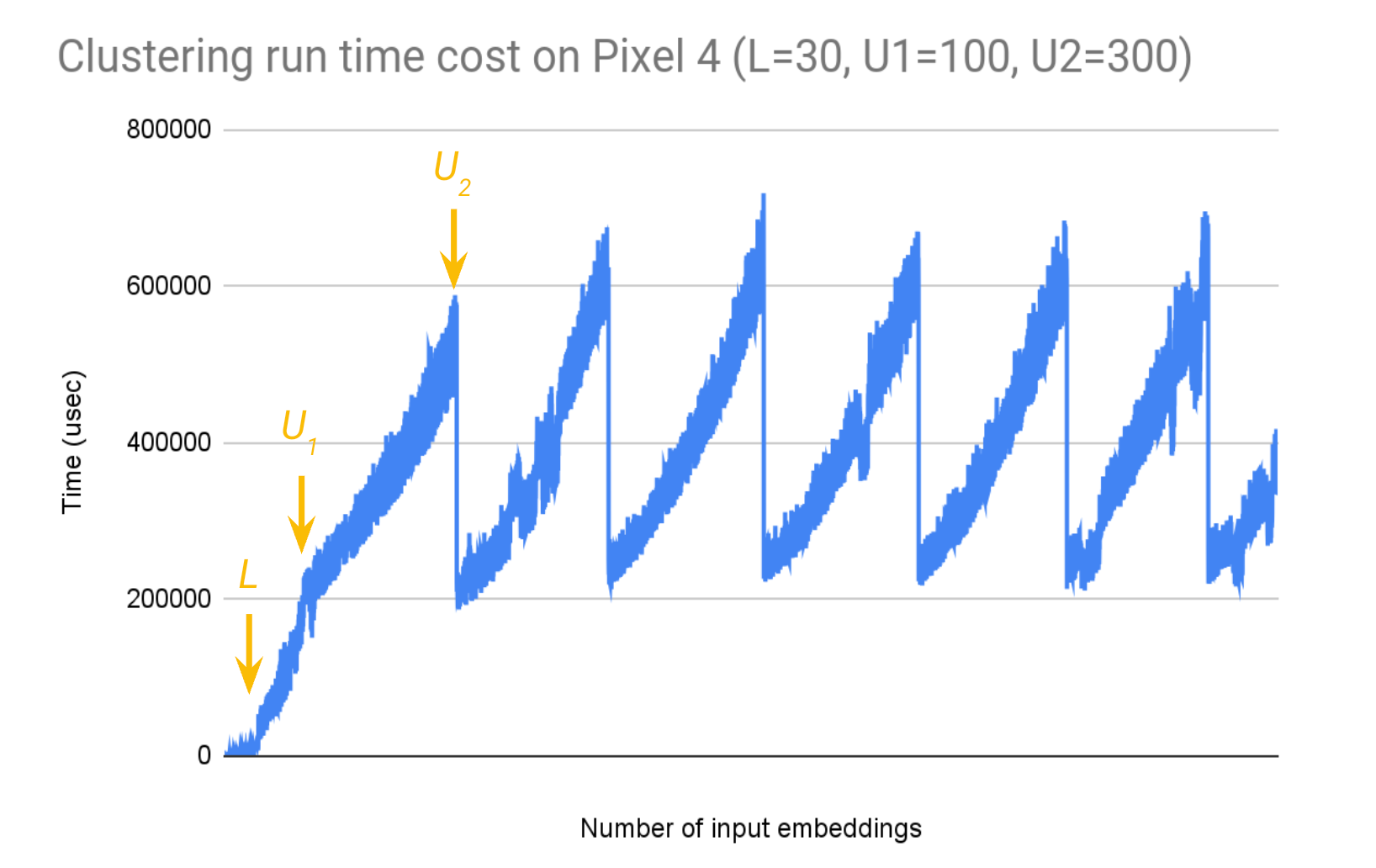}
	\vspace{-5mm}
	\caption{Plot of the multi-stage clustering runtime cost on a Pixel 4 mobile device. $L$ and $U_1$ are the lower bound and upper bound of the main clusterer, respectively; $U_2$ is the upper bound of the pre-clusterer. Roughly speaking, 100 inputs correspond to over 6 minutes, and 300 inputs correspond to approximately 20 minutes.}
	\label{fig:benchmark}
	\vspace{-5mm}
\end{figure}

To visualize the computational cost of multi-stage clustering for different values of $N$, we performed on-device benchmarking of our diarization system on a Pixel 4 mobile device, and logged the CPU runtime of each clustering step while $N$ increases during the streaming, as shown in Fig.~\ref{fig:benchmark}.

When the number of input embeddings $N$ increases, the slope is flattest in the interval $[0, L]$ corresponding to the fallback AHC clusterer. The slope is steepest in the interval $[L, U_1]$, corresponding to the main spectral clusterer. The interval $[U_1,U_2]$ is a mixture of AHC pre-clusterer and main spectral clusterer, so its slope is intermediate. Once past $U_2$, the computational cost follows a periodic pattern due to the upper bound and dynamic compression. Note that clustering generally runs slowly on mobile devices than on Linux workstations due to less available and less powerful computational resources.

\subsection{Discussions}

From the above results, we can see that an optimal value of $L$ can usually be found via evaluating the quality trade-off (both DER and speaker count errors) between short-form and long-form data using AHC and spectral clustering. However, for $U_1$ and $U_2$, there is usually no optimal value. For offline systems which have unlimited computational resources and no latency requirements, we can set both $U_1$ and $U_2$ to infinity to achieve the best long-form quality. However, for a realistic on-device application, the system must work reliably with a very limited budget of resources including CPU, memory, and battery. Thus $U_1$ and $U_2$ need to be carefully tuned based on these requirements, for a balance between computational cost and quality.

In addition, we also experimented the idea of reducing the AHC pre-clusterer computational cost with the canopy approach~\cite{mccallum2000efficient}. However, this approach is very data-dependent due to the fact that it is built upon hashing, which does not guarantee an upper bound of the computational cost, thus is infeasible for on-device applications.

We would like to highlight that the DER numbers presented in this section are not directly comparable to state-of-the-art approaches due to different assumptions and experiment setups. Specifically, we enforce various constraints such as portable model sizes and streaming latency to our setups, while state-of-the-art models usually assume unlimited resources, and do not discuss short-form performance. To the best of our knowledge, no prior work in the literature addresses the same set of challenges at the same time.

Although the evaluation results presented in this section used a specific combination of algorithms (\emph{i.e.} AHC as fallback clusterer, and spectral as main clusterer) as an example to demonstrate the effectiveness of this practical solution.We would like to point out that the general idea of the proposed multi-stage framework is not tied to using specific algorithms to specific stages. In general, the fallback cluster is designed to better detect single-vs-multiple speakers for short-form inputs; the main clusterer is designed to accurately estimate the speaker count; and the pre-clusterer is designed to enforce an upper bound to the overall computational cost. 

\section{Conclusions}
\label{sec:conclusions}
In this paper, we described a multi-stage clustering strategy for streaming on-device speaker diarization which consists of an AHC fallback clusterer, a main spectral clusterer, and an AHC pre-clusterer with dynamic compression. This strategy leverages the benefits of different unsupervised clustering algorithms. It helps us build a speaker diarization system that is capable of producing high quality results for inputs of variable lengths, with an upper bounded time and space complexity that is easy to deploy to on-device environments such as mobile phones. This system can also be easily configured to balance between computational cost and quality, in order to fit hardware constraints of different devices.

\section*{Acknowledgments}
The authors would like to thank Evan Clark, Qi Cao, Wei Xia, Hasim Sak, Alvin Zhou, Jason Pelecanos, Luiza Timariu, Allen Su, Fan Zhang, Hugh Love, Kristi Bradford, Vincent Peng, Raff Tsai, Richard Chou, Yitong Lin, Ann Lu, Kelly Tsai, Hannah Bowman, Tracy Wu, Faema Tien, Irene Lee, Taral Joglekar, Dharmesh Mokani, Ajay Dudani, Diego Melendo Casado, Nino Tasca and Alex Gruenstein.

\bibliographystyle{IEEEbib}
\bibliography{refs}

\begin{thebibliography}{10}

\bibitem{garcia2017speaker}
Daniel Garcia-Romero, David Snyder, Gregory Sell, Daniel Povey, and Alan
  McCree,
\newblock ``Speaker diarization using deep neural network embeddings,''
\newblock in {\em ICASSP}. IEEE, 2017, pp. 4930--4934.

\bibitem{dimitriadis2017developing}
Dimitrios Dimitriadis and Petr Fousek,
\newblock ``Developing on-line speaker diarization system,''
\newblock in {\em Proc. Interspeech}, 2017, pp. 2739--2743.

\bibitem{wang2017speaker}
Quan Wang, Carlton Downey, Li~Wan, Philip~Andrew Mansfield, and Ignacio
  Lopez~Moreno,
\newblock ``Speaker diarization with {LSTM},''
\newblock in {\em ICASSP}. IEEE, 2018, pp. 5239--5243.

\bibitem{park2019auto}
Tae~Jin Park, Kyu~J Han, Manoj Kumar, and Shrikanth Narayanan,
\newblock ``Auto-tuning spectral clustering for speaker diarization using
  normalized maximum eigengap,''
\newblock {\em IEEE Signal Processing Letters}, vol. 27, pp. 381--385, 2019.

\bibitem{xia2022turn}
Wei Xia, Han Lu, Quan Wang, Anshuman Tripathi, Yiling Huang, Ignacio~Lopez
  Moreno, and Hasim Sak,
\newblock ``{Turn-to-Diarize}: Online speaker diarization constrained by
  transformer transducer speaker turn detection,''
\newblock in {\em ICASSP}. IEEE, 2022, pp. 8077--8081.

\bibitem{park2021multi}
Tae~Jin Park, Manoj Kumar, and Shrikanth Narayanan,
\newblock ``Multi-scale speaker diarization with neural affinity score
  fusion,''
\newblock in {\em ICASSP}. IEEE, 2021, pp. 7173--7177.

\bibitem{zhang2019fully}
Aonan Zhang, Quan Wang, Zhenyao Zhu, John Paisley, and Chong Wang,
\newblock ``Fully supervised speaker diarization,''
\newblock in {\em ICASSP}. IEEE, 2019, pp. 6301--6305.

\bibitem{li2019discriminative}
Qiujia Li, Florian~L Kreyssig, Chao Zhang, and Philip~C Woodland,
\newblock ``Discriminative neural clustering for speaker diarisation,''
\newblock in {\em Spoken Language Technology Workshop (SLT)}. IEEE, 2021.

\bibitem{fujita2019end}
Yusuke Fujita, Naoyuki Kanda, Shota Horiguchi, Kenji Nagamatsu, and Shinji
  Watanabe,
\newblock ``End-to-end neural speaker diarization with permutation-free
  objectives,''
\newblock in {\em Proc. Interspeech}, 2019, pp. 4300--4304.

\bibitem{e2ediarizationpatent}
Quan Wang, Yash Sheth, Ignacio~Lopez Moreno, and Li~Wan,
\newblock ``Speaker diarization using an end-to-end model,'' US Patent
  US011545157B2, 2019.

\bibitem{medennikov2020target}
Ivan Medennikov et~al.,
\newblock ``Target-speaker voice activity detection: a novel approach for
  multi-speaker diarization in a dinner party scenario,''
\newblock in {\em Proc. Interspeech}, 2020.

\bibitem{ueda2022eend}
Yushi Ueda, Soumi Maiti, Shinji Watanabe, Chunlei Zhang, Meng Yu, Shi-Xiong
  Zhang, and Yong Xu,
\newblock ``{EEND-SS}: Joint end-to-end neural speaker diarization and speech
  separation for flexible number of speakers,''
\newblock {\em arXiv preprint arXiv:2203.17068}, 2022.

\bibitem{von2019all}
Thilo von Neumann, Keisuke Kinoshita, Marc Delcroix, Shoko Araki, Tomohiro
  Nakatani, and Reinhold Haeb-Umbach,
\newblock ``All-neural online source separation, counting, and diarization for
  meeting analysis,''
\newblock in {\em ICASSP}. IEEE, 2019, pp. 91--95.

\bibitem{chazan2018attention}
Shlomo~E Chazan, Sharon Gannot, and Jacob Goldberger,
\newblock ``Attention-based neural network for joint diarization and speaker
  extraction,''
\newblock in {\em 2018 16th International Workshop on Acoustic Signal
  Enhancement (IWAENC)}. IEEE, 2018, pp. 301--305.

\bibitem{park2021review}
Tae~Jin Park, Naoyuki Kanda, Dimitrios Dimitriadis, Kyu~J Han, Shinji Watanabe,
  and Shrikanth Narayanan,
\newblock ``A review of speaker diarization: Recent advances with deep
  learning,''
\newblock {\em Computer Speech \& Language}, vol. 72, pp. 101317, 2022.

\bibitem{zhang2022odysseytutorial}
Chao Zhang and Quan Wang,
\newblock ``Speaker diarization: A journey from unsupervised to supervised
  approaches,'' Odyssey: The Speaker and Language Recognition Workshop, 2022,
\newblock Tutorial session.

\bibitem{speakerlabelsblog}
Quan Wang and Fan Zhang,
\newblock ``Who said what? {Recorder's} on-device solution for labeling
  speakers,'' Google AI Blog.

\bibitem{prabhavalkar2015automatic}
Rohit Prabhavalkar, Raziel Alvarez, Carolina Parada, Preetum Nakkiran, and
  Tara~N Sainath,
\newblock ``Automatic gain control and multi-style training for robust
  small-footprint keyword spotting with deep neural networks,''
\newblock in {\em ICASSP}. IEEE, 2015, pp. 4704--4708.

\bibitem{zazo2016feature}
Rub{\'e}n Zazo~Candil, Tara~N Sainath, Gabor Simko, and Carolina Parada,
\newblock ``Feature learning with raw-waveform {CLDNNs} for voice activity
  detection,''
\newblock in {\em Proc. Interspeech}, 2016.

\bibitem{zhang2020transformer}
Qian Zhang, Han Lu, Hasim Sak, Anshuman Tripathi, Erik McDermott, Stephen Koo,
  and Shankar Kumar,
\newblock ``Transformer transducer: A streamable speech recognition model with
  transformer encoders and {RNN-T} loss,''
\newblock in {\em ICASSP}. IEEE, 2020, pp. 7829--7833.

\bibitem{tripathi2020transformer}
Anshuman Tripathi, Jaeyoung Kim, Qian Zhang, Han Lu, and Hasim Sak,
\newblock ``Transformer transducer: One model unifying streaming and
  non-streaming speech recognition,''
\newblock {\em arXiv:2010.03192}, 2020.

\bibitem{zhao2022augmenting}
Guanlong Zhao, Quan Wang, Han Lu, Yiling Huang, and Ignacio~Lopez Moreno,
\newblock ``Augmenting transformer-transducer based speaker change detection
  with token-level training loss,''
\newblock in {\em ICASSP}. IEEE, 2023.

\bibitem{cieri2004fisher}
Christopher Cieri, David Miller, and Kevin Walker,
\newblock ``The {Fisher} corpus: A resource for the next generations of
  speech-to-text,''
\newblock in {\em LREC}, 2004, vol.~4, pp. 69--71.

\bibitem{canavan1997callhome}
A~Canavan, D~Graff, and G~Zipperlen,
\newblock ``{CALLHOME American English} speech {LDC97S42},'' LDC Catalog.
  Philadelphia: Linguistic Data Consortium, 1997.

\bibitem{ge2e}
Li~Wan, Quan Wang, Alan Papir, and Ignacio~Lopez Moreno,
\newblock ``Generalized end-to-end loss for speaker verification,''
\newblock in {\em ICASSP}. IEEE, 2018, pp. 4879--4883.

\bibitem{pelecanos2021dr}
Jason Pelecanos, Quan Wang, and Ignacio~Lopez Moreno,
\newblock ``{Dr-Vectors}: Decision residual networks and an improved loss for
  speaker recognition,''
\newblock in {\em Proc. Interspeech}, 2021.

\bibitem{gulati2020conformer}
Anmol Gulati et~al.,
\newblock ``Conformer: Convolution-augmented transformer for speech
  recognition,''
\newblock in {\em Proc. Interspeech}, 2020.

\bibitem{wang2022attentive}
Quan Wang, Yang Yu, Jason Pelecanos, Yiling Huang, and Ignacio~Lopez Moreno,
\newblock ``Attentive temporal pooling for conformer-based streaming language
  identification in long-form speech,''
\newblock in {\em Odyssey: The Speaker and Language Recognition Workshop},
  2022.

\bibitem{pelecanos2022parameter}
Jason Pelecanos, Quan Wang, Yiling Huang, and Ignacio~Lopez Moreno,
\newblock ``Parameter-free attentive scoring for speaker verification,''
\newblock in {\em Odyssey: The Speaker and Language Recognition Workshop},
  2022.

\bibitem{koizumi2023miipher}
Yuma Koizumi, Heiga Zen, Shigeki Karita, Yifan Ding, Kohei Yatabe, Nobuyuki
  Morioka, Yu~Zhang, Wei Han, Ankur Bapna, and Michiel Bacchiani,
\newblock ``Miipher: A robust speech restoration model integrating
  self-supervised speech and text representations,''
\newblock {\em arXiv preprint arXiv:2303.01664}, 2023.

\bibitem{chojnacka2021speakerstew}
Roza Chojnacka, Jason Pelecanos, Quan Wang, and Ignacio~Lopez Moreno,
\newblock ``{SpeakerStew}: Scaling to many languages with a triaged
  multilingual text-dependent and text-independent speaker verification
  system,''
\newblock in {\em Proc. Interspeech}, 2021.

\bibitem{librivox_2020_1}
LibriVox,
\newblock ``{L}ibri{V}ox: {F}ree public domain audiobooks,''
  https://librivox.org/, 2020.

\bibitem{fan_2020_1}
Y.~Fan, J.W. Kang, L.T. Li, K.C. Li, H.L. Chen, S.T. Cheng, P.Y. Zhang, Z.Y.
  Zhou, Y.Q. Cai, and D.~Wang,
\newblock ``{CN-Celeb: A} challenging {C}hinese speaker recognition dataset,''
\newblock in {\em ICASSP}. IEEE, 2020.

\bibitem{Cieri2004_1}
Christopher Cieri, David Graff, Owen Kimball, Dave Miller, and Kevin Walker,
\newblock ``Fisher {E}nglish training speech parts 1 and 2, {LDC2004S13,
  LDC2005S13},''
\newblock {\em Web Download. Philadelphia: Linguistic Data Consortium}, 2004.

\bibitem{Brandschain2020_1}
Linda Brandschain, Kevin Walker, David Graff, Christopher Cieri, Abby Neely,
  Nikki Mirghafori, Barbara Peskin, Jack Godfrey, Stephanie Strassel, Fred
  Goodman, George~R. Doddington, and Mike King,
\newblock ``Mixer 4 and 5 speech {LDC2020S03},''
\newblock {\em Web Download. Philadelphia: Linguistic Data Consortium}, 2020.

\bibitem{garofolo_1993_1}
John~S. Garofolo, Lori~F. Lamel, William~M. Fisher, Jonathan~G. Fiscus,
  David~S. Pallett, Nancy~L. Dahlgren, and Victor Zue,
\newblock ``{TIMIT} acoustic-phonetic continuous speech corpus {LDC93S1},''
\newblock {\em Web Download. Philadelphia: Linguistic Data Consortium}, 1993.

\bibitem{lippmann1987multi}
Richard Lippmann, Edward Martin, and D~Paul,
\newblock ``Multi-style training for robust isolated-word speech recognition,''
\newblock in {\em ICASSP}. IEEE, 1987, vol.~12, pp. 705--708.

\bibitem{ko2017study}
Tom Ko, Vijayaditya Peddinti, Daniel Povey, Michael~L Seltzer, and Sanjeev
  Khudanpur,
\newblock ``A study on data augmentation of reverberant speech for robust
  speech recognition,''
\newblock in {\em ICASSP}. IEEE, 2017, pp. 5220--5224.

\bibitem{kim2017generation}
Chanwoo Kim, Ananya Misra, Kean Chin, Thad Hughes, Arun Narayanan, Tara
  Sainath, and Michiel Bacchiani,
\newblock ``Generation of large-scale simulated utterances in virtual rooms to
  train deep-neural networks for far-field speech recognition in {Google
  Home},''
\newblock in {\em Proc. Interspeech}, 2017.

\bibitem{garcia-romero_2012_1}
Daniel Garcia-Romero, Xinhui Zhou, and Carol~Y. Espy-Wilson,
\newblock ``Multicondition training of {G}aussian {PLDA} models in i-vector
  space for noise and reverberation robust speaker recognition,''
\newblock in {\em ICASSP}. IEEE, 2012, pp. 4257--4260.

\bibitem{lei_2012_1}
Yun Lei, Lukas Burget, Luciana Ferrer, Martin Graciarena, and Nicolas Scheffer,
\newblock ``Towards noise-robust speaker recognition using probabilistic linear
  discriminant analysis,''
\newblock in {\em ICASSP}. IEEE, 2012, pp. 4253--4256.

\bibitem{avila_2014_1}
Anderson~R. Avila, Milton Sarria-Paja, Francisco~J. Fraga, Douglas
  O'Shaughnessy, and Tiago~H. Falk,
\newblock ``Improving the performance of far-field speaker verification using
  multi-condition training: The case of {GMM-UBM} and i-vector systems,''
\newblock in {\em Proc. Interspeech}, 2014, pp. 1096--1100.

\bibitem{snyder_2018_1}
David Snyder, Daniel Garcia-Romero, Gregory Sell, Daniel Povey, and Sanjeev
  Khudanpur,
\newblock ``X-vectors: Robust {DNN} embeddings for speaker recognition,''
\newblock in {\em ICASSP}. IEEE, 2018.

\bibitem{huang_2019_1}
Chien-Lin Huang,
\newblock ``Exploring effective data augmentation with {TDNN-LSTM} neural
  network embedding for speaker recognition,''
\newblock in {\em IEEE Automatic Speech Recognition and Understanding Workshop
  (ASRU)}, 2019.

\bibitem{pan1999complexity}
Victor~Y Pan and Zhao~Q Chen,
\newblock ``The complexity of the matrix eigenproblem,''
\newblock in {\em Proceedings of the thirty-first annual ACM symposium on
  Theory of computing}, 1999, pp. 507--516.

\bibitem{coppersmith1987matrix}
Don Coppersmith and Shmuel Winograd,
\newblock ``Matrix multiplication via arithmetic progressions,''
\newblock in {\em Proceedings of the nineteenth annual ACM symposium on Theory
  of computing}, 1987, pp. 1--6.

\bibitem{dean2008mapreduce}
Jeffrey Dean and Sanjay Ghemawat,
\newblock ``{MapReduce}: simplified data processing on large clusters,''
\newblock {\em Communications of the ACM}, vol. 51, no. 1, pp. 107--113, 2008.

\bibitem{bredin2017pyannote}
Herv{\'e} Bredin,
\newblock ``\texttt{pyannote.metrics}: a toolkit for reproducible evaluation,
  diagnostic, and error analysis of speaker diarization systems,''
\newblock in {\em Proc. Interspeech}, 2017, pp. 3587--3591.

\bibitem{carletta2005ami}
Jean Carletta et~al.,
\newblock ``The {AMI} meeting corpus: A pre-announcement,''
\newblock in {\em International workshop on machine learning for multimodal
  interaction}. Springer, 2005, pp. 28--39.

\bibitem{ryant2018first}
Neville Ryant, Kenneth Church, Christopher Cieri, Alejandrina Cristia, Jun Du,
  Sriram Ganapathy, and Mark Liberman,
\newblock ``First {DIHARD} challenge evaluation plan,''
\newblock {\em 2018, Tech. Rep.}, 2018.

\bibitem{janin2003icsi}
Adam Janin et~al.,
\newblock ``The {ICSI} meeting corpus,''
\newblock in {\em ICASSP}. IEEE, 2003, vol.~1, pp. I--I.

\bibitem{mccallum2000efficient}
Andrew McCallum, Kamal Nigam, and Lyle~H Ungar,
\newblock ``Efficient clustering of high-dimensional data sets with application
  to reference matching,''
\newblock in {\em ICKDDM}. ACM, 2000, pp. 169--178.

\end{thebibliography}

\end{document}